\magnification=\magstep1
\hoffset.5truein
\hsize 15.0true cm
\vsize 22.0true cm
\pageno=1

\noindent{\bf Pentaquark in a supersmmetric quark-diquark model}
\bigskip\bigskip
\hskip 1cm {\bf D.B. Lichtenberg}
\medskip
\hskip 1cm Physics Department, Indiana University, Bloomington, IN 47405, USA
\medskip
\hskip 1cm E-mail: lichten@indiana.edu
\bigskip

\bigskip
\hskip 1cm {\bf Abstract}

\hskip 1cm According to QCD, there exists a broken dynamical supersymmetry
between an antiquark and a diquark. This supersymmetry can be used to
relate the mass of a pentaquark to the mass of an antibaryon
by replacing two antiquarks in an antibaryon by two diquarks to 
form a pentaquark. Using this technique, we find that the mass of an exotic
pentaquark with strangeness plus 1 is greater than 1.74 GeV, or at least
200 MeV larger than that of the reported $\Theta^+$ pentaquark. Furthermore,
there is no reason for the pentaquark to be narrow; on the contrary,
it is expected to be so broad that it will be difficult to observe. 
\bigskip\bigskip
\noindent Several groups [1--4] have claimed evidence for a
baryon called $\Theta^+(1540)$ with positive strangeness. Additional
observations were reported at a recent conference on Quarks and Nuclear
Physics [5]. A baryon with strangeness 
$S=1$ cannot consist of three quarks, but must contain at least
four quarks and an antiquark, in other words, must be a pentaquark
or a still more complicated object. Other 
people [5, 6] have cast doubt on whether the pentaquark has in fact
been established. 

Of course it is an experimental question whether the $\Theta^+$ exists
or not, but it is still of interest to see what QCD has to say on the
subject. So far, to our knowledge, calculations of the pentaquark
mass have  been made in lattice QCD only in the quenched approximation, 
We mention two such calculations, one by Sasaki [7], who finds the
lowest-mass baryon with strangeness $S=1$ to be more than 200 MeV higher than the
reported mass of the $\Theta^+$ (he says not to take the predicted mass too
seriously).  On the other hand, Csikor et al [8]
calculate a mass that they say is consistent with that of the $\Theta^+$.
In the absence of a unique value of the $\Theta^+$ mass from
lattice calculations, we use a simple model based on QCD to predict
a lower limit to the mass of the $S=1$ pentaquark. We use the framework
of the constituent quark model.

According to QCD, an approximate dynamical supersymmetry exists
between an antiquark and a diquark. The first
person to point out  a supersymmetry between antiquarks (half-integral spin)
and diquarks (integral spin) was Miyazawa [9] in 1966, several years
before it was suggested that a supersymmetry might exist between
elementary quarks and squarks, between electrons and selectrons, etc.
In fact, we think that Miyazawa was the first person to apply 
the idea of supersymmetry to particle physics. (Miyazawa did not
use either the word ``diquark'' or the word ``supersymmetry'' in
his paper, but the concepts are there.)

The point is that both a diquark and an antiquark belong to an
antitriplet of SU(3)-color, and to first approximation the 
interaction of QCD depends only on color. A diquark can also belong
to a sextet of color, but a sextet diquark is irrelevant to 
the supersymmetry we are discussing. In lowest-order QCD perturbation
theory, two quarks in a color sextet are subject to a repulsive force,
with the consequence that a sextet diquark has a higher mass than
an antitriplet diquark. We have no reason to believe that this
result will change in full QCD.

Catto and G\"ursey [10]
were the first to point out that the supersymmetry between 
antiquark and antitriplet diquark
is an approximate consequence of QCD and that the supersymmetry
is responsible for the fact that mesons and baryons have 
Regge trajectories with approximately the same slope. 
A review of diquarks, including a section on supersymmetry, appeared
in a paper in 1993 [11].

We can think of at least three things that break the supersymmetry:

1) A quark and a diquark have different masses. It is easy to 
take this mass difference into account in approximate fashion
by assigning a diquark to
have a mass equal to the sum of the masses of the quarks it contains. 

2) A quark and a diquark have different sizes. A diquark may be
almost the size of a hadron, whereas a current quark is assumed
to be pointlike. However, a constituent quark, with its
cloud of gluons and quark-antiquark pairs, is certainly not pointlike.
We neglect the size difference between a constituent antiquark and a diquark.

3) A quark and a diquark have different spins, and in QCD there
are spin-dependent interactions. In the present paper, we adopt
a procedure which does not quantitatively take into account the spin
dependence of quark and diquark interactions. However, we deduce 
from the systematics of baryon masses that a spin-zero diquark
has a lower mass than a diquark of spin one. 

A few years ago, similar ideas were used [12] to predict the masses
of exotic mesons (four-quark states) and dibaryons (six-quark states)
but not the properties of pentaquarks.  It was found that quadriquarks
and sexaquarks,
with the possible exception of states containing two $b$ quarks, 
are highly unbound in the model and probably have decay widths too
large to be easily observed.

Here, we do not follow Ref. [12] in detail but use simpler arguments,
which we hope are more general, to show that a pentaquark state with
properties of the $\Theta^+$ does not exist in the model. 
We can choose quark constituent masses 
such that a ground-state hadron is the sum of the masses of the quarks
it contains. Except in the heavy-quark limit, such a procedure 
requires us to assign different masses to quarks of the same flavor
in different hadrons. But if we are interested just in lower limits
to certain hadron masses, we can avoid this complication. 

First, we consider an example of a meson and a baryon containing only $u$ and
$d$ quarks, and neglect the mass difference between the $u$ and $d$.
The lightest meson containing these quarks is the pion,
while the lightest baryon is the nucleon. If the $u$ and $d$ quarks were
infinitely heavy, the nucleon mass would be 3/2 times the mass of the pion,
but of course actually the nucleon mass is considerably larger than 3/2 times 
the pion mass. Part of the reason is that the pion is anomalously light, but
the result seems to be more general. For example, if we replace
a $\bar u$ antiquark by a light diquark in a $\bar K$ meson, 
we obtain a strange baryon.
The lightest strange baryon is the $\Lambda$, and its mass
is larger than the $K$ mass by an amount that is greater than the
extra mass of a light constituent quark. The additional mass acquired
by a hadron when we replace an antiquark by a diquark is, of course,
a breaking of the dynamical supersymmetry. The rule seems to be
that {\it when we replace an antiquark by a diquark in a meson, the resulting
baryon has a mass that is greater than the meson mass by an amount
that is larger than the mass of the additional quark}. 

Likewise, we expect that if we replace two antiquarks in an antibaryon
by diquarks, the mass of the resulting baryon  (a pentaquark) will
be larger than the original antibaryon mass by an amount greater
than the two extra quark masses. We can think of no reason why the
systematics here should be any different from the meson--baryon case. 
A model in which a pentaquark consists of two diquarks and an antiquark
has previously been considered by Jaffe and Wilczek [13], but not within
the framework of diquark-antiquark supersymmetry.

Without further ado, we turn to the problem at hand. Using the
fact that the nucleon has a mass of 939 MeV and that it contains
three constituent quarks, we choose the mass of a $u$ or $d$
quark to be 313 MeV. (We could of course choose a somewhat
different constituent quark mass at the expense of including an
appropriate binding energy.) The lightest baryon containing a strange 
quark is the $\Lambda$, with a mass of 1115 MeV. If we consider an 
anti-$\Lambda$ and replace its $\bar u$ and $\bar d$  antiquarks by light 
diquarks, we add $2\times 313= 626$ MeV to the mass of the $\Lambda$ (1115 MeV) 
to obtain a lower limit on the mass of a pentaquark with strangeness
$S= +1$. We obtain
$$M(\Theta^+)> 1741 \quad {\rm MeV}.$$
This value is at least 200 MeV larger than the reported mass [1--5]
of the $\Theta^+$ baryon. 

Because the two spin-zero $ud$ diquarks in the pentaquark are identical bosons,
they cannot have a symmetric spatial wave function. We see this as follows:
The wave function of the two diquarks is antisymmetic
in color. Therefore, their space wave function must also be antisymmetric
so as to make an overall symmetric wave function. 
One way to accomplish this is by putting the two diquarks in a state
with orbital angular momentum one. In this state the energy will probably
be a few hundred MeV higher than 1740 MeV, and the parity will be positive.
 
If, in the model, one of the diquarks has spin one
then, by symmetry, that diquark must also have isospin one, in 
contradiction to the claim that the pentaquark has isospin zero. 
If both diquarks have spin and isospin one, the total isospin 
can be zero. From two spin-one diquarks we can construct an 
antisymmetric state with spin one. This spin can combine with the
spin of the $\bar s$ quark to form total spin 1/2 or 3/2, but the
systematics of spin-dependent forces tells us that the state with
spin 1/2 will lie lower.  In this case
the two diquarks have a symmetric spatial wave function and 
negative parity, but again we estimate that the energy will be
a few hundred MeV higher than 1740 MeV because a spin-one diquark
is heavier than a diquark of spin zero. It is not clear whether
the state with negative or positive parity will lie lower. 

In the model there is nothing to prevent a rapid decay of
the pentaquark by color rearrangement into nucleon + kaon. Because the
$Q$-value of the decay is expected to be well over 300 MeV, the
pentaquark should have such a broad width as to make it hard to observe
as a resonance, even in the case with orbital angular momentum zero.  

We conclude that in a supersymmetric quark-diquark model of hadrons
which is motivated by QCD, we do not expect the existence 
of a state with the reported properties of the $\Theta^+$ baryon. 
\bigskip\bigskip

\noindent{\bf Acknowledgments}
\bigskip
I am grateful to Steven Gottlieb for information about lattice calculations
of the pentaquark mass. A preliminary version of this paper was presented
as a talk at the International Conference on Quarks and Nuclear Physics [5], 

\bigskip\bigskip
\noindent{\bf References}
\bigskip
[1] T. Nakano et al., Phys. Rev. Lett. {\bf 91}, 012002 (2003).

[2] S. Stepanyan et al., Phys. Rev. Lett. {\bf 91}, 252001 (2003).

[3] J. Barth et al., Phys. Lett. B {\bf 572}, 127 (2003).

[4] V.V. Barmin et al., Phys. At. Nucl. {\bf 66}, 1715 (2003).

[5] International Conference on Quarks and Nuclear Physics,
May 23--28, 2004, Bloomington, Indiana.

[6] A.R. Dzierba et al., Phys. Rev. D {\bf 69}, 051901R (2004).

[7] S. Sasaki, hep-lat/0310014 (2003).

[8] F. Csikor et al., hep-lat/0309090. JHEP 0311, 070 (2003).

[9] H. Miyazawa, Prog. Theor. Phys. {\bf 36}, 1266 (1966).

[10] S. Catto and F. G\"ursey, Nuovo Cimento {\bf 86}, 201 (1985).

[11] M. Anselmino et al., Rev. Mod. Phys. {\bf 65} 1199 (1993) 

[12] D.B. Lichtenberg, R. Roncaglia, and E. Predazzi, J. Phys. G
Nucl. Part. Phys. {\bf 23}, 865 (1997).

[13] R. Jaffe and F. Wilczek, Phys. Rev. Lett. {\bf 91}, 232003 (2003).
\bye